\documentclass[11pt]{amsart}
\usepackage{graphics}
\usepackage{hyperref}
\numberwithin{equation}{section}
\def \dd{{\rm d}}

\begin{document}

\title[]{The physical meaning of the ``boost-rotation symmetric''
solutions within the general interpretation of Einstein's theory of gravitation}%
\author{Salvatore Antoci}%
\address{Dipartimento di Fisica ``A. Volta'' and CNR, Pavia, Italy}%
\email{Antoci@fisicavolta.unipv.it}%
\author{Dierck-Ekkehard  Liebscher}%
\address{Astrophysikalisches Institut Potsdam, Potsdam, Germany}%
\email{deliebscher@aip.de}%
\author{Luigi Mihich}%
\address{Dipartimento di Fisica ``A. Volta'', Pavia, Italy}%
\email{Mihich@fisicavolta.unipv.it}%
\begin{abstract}
The answer to the question, what physical meaning should be attributed to
the so-called boost-rotation symmetric exact solutions to
the field equations of general relativity, is provided within
the general interpretation scheme for the ``theories of relativity'',
based on group theoretical arguments, and set forth by Erich Kretschmann
already in the year 1917.
\end{abstract}
\maketitle
\section{Introduction}
In the same year 1915, when Einstein and Hilbert
\cite{Einstein1915,Hilbert1915} gave the final mathematical
expression to the long efforts done by Einstein for finding a generally
covariant theory of gravitation based on the absolute differential
calculus of Ricci and Levi Civita \cite{RLC1900}, Erich Kretschmann
published a long article \cite{Kretschmann1915}, entitled
``\"Uber die prinzipielle Bestimmbarkeit der berechtigten
Bezugssysteme beliebiger Relativit\"atstheorien'', in which a
minute analysis of the relation between observation and mathematical
structure in a theory possessing a generic postulate of relativity
is developed. No wonder then, if two years later, with the paper
\cite{Kretschmann1917} entitled ``\"Uber den physikalischen Sinn
der Relativit\"atspostulate; A. Einsteins neue und seine urspr\"ungliche
Relativit\"atstheorie'', the same author produced an analysis of the
relation between the ``special'' and the ``general'' theory of
relativity that had to become a source of permanent enlightenment
for the relativists. The analysis relies on a fundamental distinction
between the group of invariance and the group of covariance of a theory,
that appears to have escaped the attention both of Einstein and of Hilbert.
A faithful account of Kretschmann's result was given by Philipp Frank's
review \cite{Frank1917} of the paper, that reads, in English translation:\par
``Einstein understands, under his general principle of relativity,
the injunction that the laws of nature must be expressed through
equations that are covariant with respect to arbitrary coordinate
transformations. The Author shows now that any natural phenomenon
obeying any law can be described by generally covariant equations.
Therefore the existence of such equations does not express any physical
property. For instance the uniform propagation of light in a space free from
gravitation can be expressed also in a covariant way. However, there is a
representation of the same phenomena that admits only a more restricted group
(the Lorentz transformations). This group, that cannot be further restricted
by any representation of the phenomena, is characteristic of the
system under question. The invariance with respect to it is a physical property
of the system and, in the sense of the Author, it represents the postulate of
relativity for the corresponding domain of phenomena.\par
In Einstein's general theory of relativity, through appropriate choice of the
coordinates, the field equations can be converted in a form that is no longer
covariant under the group of coordinate transformations. The Author provides a
series of examples of such conversions. But the equations converted in this way
in general no longer admit any group, and in this sense Einstein's theory of
general relativity is an ``absolute theory'', while the special theory of
relativity satisfies the postulate of relativity for the Lorentz transformations
also in the sense of the Author.''\par
Kretschmann's viewpoint, that deprives the coordinates and the
covariance under general transformations of physical meaning in a nearly complete
way\footnote{A residual physical meaning is however left. In facts Kretschmann
embraces Einstein's view \cite{Einstein1916} that the description of the
whole physical experience can be reduced to accounting for spatiotemporal
coincidences. Therefore coordinates have no physical meaning in themselves,
but of course a restriction of physical origin on the admissible coordinate
transformations is mandatory: since a coordinate system must faithfully
absolve the physical function of reckoning the spacetime coincidences,
it must preserve the individuality of the single event. To this end, only one to
one coordinate transformations can be allowed for.},
was recognised correct by Einstein \cite{Einstein1918}, and has become
part and parcel of the present day understanding of ``general relativity'':
coordinates, and the values that the components of tensorial entities
may assume with respect to a given chart, do not matter; the objective
physical content of the theory is written in the geometry of the
manifold, and it can be read only through the invariant quantities
associated with the latter. The same acceptance was met with by Kretschmann's
way of assessing the ``relativity content'' of a given theory.
For him, it should not be ascertained through the group of covariance
allowed by the particular expression adopted for writing the equations
of that theory, but through its group of invariance, meant to be
``a physical property of the system'', directly inscribed by the Killing
vectors in the intrinsic, geometric structure of the manifold.\par
    Kretschmann's analysis \cite{Kretschmann1917}, however, only
considered the group of invariance of a general solution to the field
equations of ``general relativity'', that contains only the identity,
and the group of invariance for the particular solution of the same theory
that occurs when $R_{iklm}=0$, namely, the inhomogeneous Lorentz group.
Since that time, solutions of Einstein's theory of 1915 whose
groups of invariance correspond to a ``relativity content'' intermediate
between the above mentioned extremes have been found, and investigated
at length by the relativists. To these solutions belong the so called
``boost rotation symmetric'' solutions. Scope of the present paper
is the assessment of the physical meaning of these solutions
as dictated, in keeping with Kretschmann's idea, by the geometric
structure of their manifolds.
\section{The``boost-rotation symmetric'' solutions}
The perusal of the literature dealing with the
``boost-rotation symmetric'' solutions, spanning a
time interval of four decades, shows that all the
vacuum solutions associated with nonspinning sources
\cite{Bondi1957}, \cite{BS1964}, \cite{IK1964},
\cite{Bonnor1966}, \cite{Bicak1968,BHS1983,BS1984},
\cite{Bonnor1983,Bonnor1988}, \cite{BS1989}, (see also \cite{SKMHH2003}) can be generated
in one and the same way by starting from some solution
belonging to the class found long ago by Weyl \cite{Weyl1917}
and by Levi-Civita \cite{Levi-Civita1919}. For the convenience of the reader,
the definition of the latter class of solutions in
the canonical coordinates introduced by Weyl is reported in
Appendix \ref{A}. For instance, the solution like the one reported
in \cite{Bonnor1988} can be obtained by choosing the function
$\psi$, that fulfils the ``potential'' equation (\ref{A2}), in such a way that
\begin{equation}\label{2.1}
\psi={\frac 12}\ln\left[(r^2+z^2)^{\frac 12}+z\right]
+{\frac 12}\ln\frac{r_1+r_2-2l}{r_1+r_2+2l},
\end{equation}
where $r_i=[(z-z_i)^2+r^2]^{\frac 12}$, and the positive constants
$z_1$ and $z_2$ are so chosen that $z_2-z_1=2l> 0$. If $\psi$ were
a Newtonian potential, its particular expression (\ref{2.1})
would correspond to the sum of the potential of two rods
both endowed with linear mass density $\sigma=1/2$, and lying
on the $z$-axis. One of the rods extends itself from $z=0$ to $z=-\infty$,
while the other one covers the segment between $z_1$ and $z_2$.
But this imagery is just a ``Bildraum'' deception for, if only the
semi-infinite rod were present, the metric generated by the Weyl
method would be such that $R_{iklm}=0$, while, if only the finite
rod with $z_1<z<z_2$ were present, the solution would be in one to one
correspondence with the original \cite{Schwarzschild1916}
Schwarzschild solution\footnote{whose manifold, at variance with the
``Schwarzschild'' solution referred to in the literature, that was actually
proposed by Hilbert \cite{Hilbert1917}, does not cover the ``inner region''
of the latter.} for a mass $m=l$.\par
In order to obtain the ``boost-rotation symmetric'' solution corresponding to
this Weyl field, one goes over to the primed cylindrical polar coordinates
${x'}^1=z'$, ${x'}^2=r'$, ${x'}^3=\varphi'$, ${x'}^4=t'$ from the unprimed,
canonical coordinates specified in Appendix \ref{A}, through the coordinate
transformation
\begin{eqnarray}\label{2.2}
z'=\pm[(r^2+z^2)^{\frac12}+z]^{\frac12}\cosh t,\\\label{2.3}
r'=[(r^2+z^2)^{\frac12}-z]^{\frac12},\\\label{2.4}
t'=[(r^2+z^2)^{\frac12}+z]^{\frac12}\sinh t,\\\label{2.5}
\varphi'=\varphi.
\end{eqnarray}
We note in passing that this transformation neither conforms to
Einstein's mentioned injunction that coordinate transformations
should be one-to-one, in order to preserve the identity of the events,
nor obeys the prescriptions by Hilbert and Lichnerowicz about the admissible
transformations of coordinates \cite{Hilbert1917,Lichnerowicz1955}.
In fact, besides the obvious doubling of the Weyl manifold due to the
$\pm$ sign of (\ref{2.2}), one notes that the events of the original
manifold for which $t$ is finite and otherwise arbitrary, $r=0$, $-\infty<z<0$,
in the primed coordinates all end up in the coordinate plane for which $z'=t'=0$
in a way that only depends on $z$, but not on $t$. Therefore the transformation
of Eqs. (\ref{2.2})-(\ref{2.5}) loses track of the individuality of events
as it is specified within the Weyl manifold.
\begin{figure}[ht]
\includegraphics{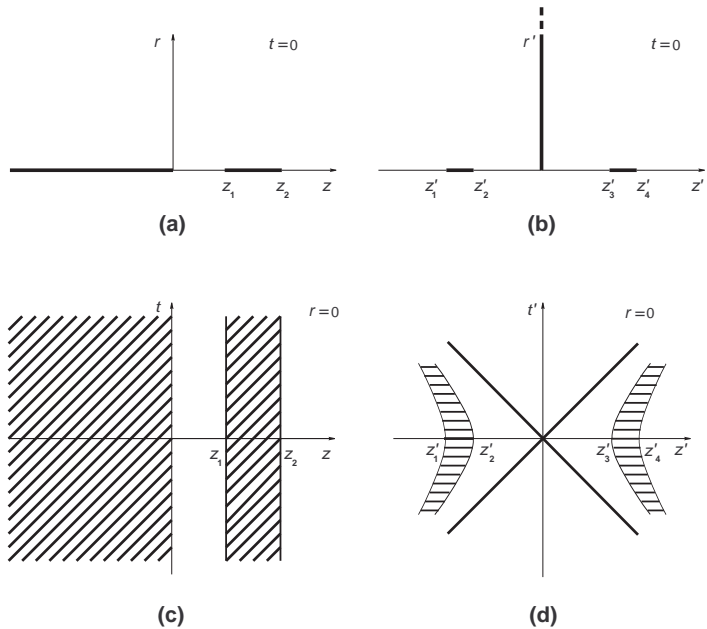}
\caption{Left side: sketch, in canonical coordinates, of the ``Newtonian
sources'' corresponding to the ``potential'' $\psi$ of Eq. (\ref{2.1}).
a) $z$, $r$ diagram for $t=0$. c) $z$, $t$ diagram for $r=0$.
Right side: representation in the primed coordinates corresponding
to the transformation of Eqs. (\ref{2.2})-(\ref{2.5}).
b) $z'$, $r'$ diagram for $t=0$. d) $z'$, $t'$ diagram for $r=0$.}
\end{figure}
A glance to the drawings (a) and (b) of Figure 1 shows that the semi-infinite
rod should go in the entire plane $z'=0$, while the finite rod
$z_1z_2$ is doubled in the mirror images ${z'}_1{z'}_2$ and
${z'}_3{z'}_4$. This is true for $t=0$, and also for any value
of $t$, but with a different scale along $z'$. When $t$ is finite,
the shaded area in the left part of (c) goes in the origin
of the $z'$, $t'$ diagram (d), while, when $t=\pm\infty$, it
is spread on the diagonals of the same diagram. The strip on the
right part of (c) goes instead in the two shaded areas of (d),
delimited by two hyperbolae that, in the primed representation,
asymptotically approach the two diagonals. Diagram (d) shows how
the transformation (\ref{2.2})-(\ref{2.5}) produces a doubling in the
representation, since the whole $z$, $t$ plane, cut along the negative
part of the $z$ axis, appears twice in the left and right
quadrants of (d), in a way akin to the duplication of the original Schwarzschild
manifold in the left and right quadrants of the Kruskal
manifold \cite{Synge1950,Kruskal1960,Szekeres1960}. Moreover, like in the
latter case, the interval, when written in the primed
coordinates, happens to be a solution of Einstein's equations not only
within the left and right quadrants, but also in the upper and lower
ones, i.e. the full diagram yields both a duplication and an extension
of the original Weyl manifold. It is remarkable that the solution of the
upper and of the lower quadrants could be obtained \cite{Beck1925} also from
a Weyl solution, by first subjecting it to the formal change
\begin{equation}\label{2.6}
t\rightarrow iz, \ z\rightarrow it, \ \ i=\sqrt{-1},
\end{equation}
that preserves the reality of the interval.

\section{A matter of interpretation}
When confronted with the diagrams of the left and of the right
sides of Figure 1, one is awestruck by the mathematical beauty
of the coordinate transformation that has brought the two standing
rods of the Weyl solution, apparently two standing masses that no doubt need a
strut to be held forever at rest despite their mutual gravitational
pull \cite{Weyl1919,BW1922}, into a bifurcate horizon and two masses
executing hyperbolic motions independent of each other, of course thanks
to struts providing the necessary push. This transformation is not
a peculiarity that only applies to the Weyl solution defined by
Eq. (\ref{2.1}); it provides the cornerstone upon which all the
``boost-rotation symmetric'' solutions of Refs. \cite{BS1964}-\cite{BS1989}
are built. However, both the static character of the originating Weyl metrics,
and the particular time dependent behaviour in the left and right quadrants
seen in the primed coordinates of diagram (d), are just a coordinate imagery,
possibly a ``Bildraum'' deception, because, as taught long ago \cite{Kretschmann1917}
by Kretschmann and Einstein \cite{Einstein1916}, since the coordinates are
nearly devoid of physical meaning, such is also the case for the expressions
that a solution takes in a certain chart. We have to search for the physical
meaning of a solution by studying its invariant features, in particular its
group of invariance.\par
The Weyl-Levi Civita solutions are particular examples, endowed with axial
symmetry, of the general class of static solutions. It is generally said in the
textbooks that these solutions are invariantly defined by the
existence of a timelike Killing vector $\xi^i$ that is also
hypersurface orthogonal:
\begin{equation}\label{3.1}
\xi_i\xi^i>0, \ \ \xi_{i;k}+\xi_{k;i}=0, \ \ \xi_{[i}\xi_{k,l]}=0.
\end{equation}
It must be noticed, however, that this definition is not stringent enough:
a manifold for which $R_{iklm}=0$ of course possesses a vector that fulfils
(\ref{3.1}), because, since the group of invariance of that manifold is
the inhomogeneous Lorentz group of special relativity, it possesses an infinity
of them. But, when  $R_{iklm}\ne 0$, it generally happens that at each event
equations (\ref{3.1}) allow only for a unique way of defining the direction
of the timelike, hypersurface orthogonal Killing vector. This uniqueness is crucial
for the physical interpretation. When it occurs, the ``relativity content'' of
the manifold is the following: the Killing vectors
fulfilling (\ref{3.1}) provide a one parameter group of invariance,
and their hypersurface orthogonality yields a unique, intrinsic, absolute distinction
between space and time, namely, provides a gravitational aether in which
absolute space, absolute time, absolute rest are meaningful physical
notions, since they are invariantly inscribed in the geometry of the manifold.
In general relativity, only solutions endowed with this intrinsic structure
can be properly named static. Weyl-Levi Civita solutions with a nonvanishing
Riemann tensor are static in the sense defined above; the manifolds associated
to them possess however a further symmetry, since their group of invariance is
constituted by the two Killing vectors that define respectively the translation
along absolute time and the spatial rotation around a given axis.\par
As a consequence, the physical reading of diagrams (a) and (c), and the physical
reading of (b) and of the left and right quadrants of (d) cannot be but one and
the same: in an absolute, invariant sense, we have to do with bodies
at rest with respect to the manifold; despite their mutual gravitational
pull, they are kept in such a condition by the existence of a well
investigated  \cite{Weyl1919,BW1922} strut between them.\par

\section{The bifurcate horizon is singular in an invariant sense}
One can object that, although the left and right quadrants of diagram (d)
are no doubt static in the absolute sense explained above, hence
cannot provide an idealised model for the process of emission and absorption
of gravitational radiation by material bodies, the upper and lower
quadrants are indeed time dependent in an absolute sense. In fact, on crossing
the horizon by going from the left and right quadrants to the upper
and the lower ones, the hypersurface orthogonal, timelike Killing
vector becomes null and then spacelike. The upper and lower quadrants
provide in fact two distorted copies of a time dependent solution endowed with
cylindrical symmetry belonging to the class that Beck
found \cite{Beck1925} in 1925 from the Weyl-Levi Civita solutions through
the formal change (\ref{2.6}), and their intrinsic reading
is completely different from the one that applies to the left and right
quadrants.\par
It has been remarked above that the extension of diagram (d) is
reminiscent of the duplication and extension of the original Schwarzschild
solution that goes under the name of
Kruskal \cite{Synge1950,Kruskal1960,Szekeres1960}; for that extension,
it has been proved already \cite{AL2001,ALM2003} that a local, invariant, intrinsic
singularity occurs when approaching the horizon. The same thing occurs
with the extension of diagram (d). Since the singularity is defined in
an invariant way, its existence can be conveniently ascertained by
using Weyl's canonical coordinates.\par
The quantity under question is the norm $\alpha$ of the
four acceleration
\begin{equation}\label{4.1}
a^i\equiv\frac{\dd u^i}{\dd s}+\Gamma^i_{kl}u^ku^l
\end{equation}
of a test particle whose world line is a line of absolute
rest in the above explained sense. Besides being invariant, this
quantity is intrinsic to the manifold, like it is the world line
of absolute rest.\par
The nonvanishing components of $a^i$ for a worldline of
rest in a Weyl-Levi Civita solution are
\begin{equation}\label{4.2}
a^1=\exp{(2\psi-2\gamma)}\frac{\partial\psi}{\partial z},\;
a^2=\exp{(2\psi-2\gamma)}\frac{\partial\psi}{\partial r}.
\end{equation}
When calculating $\alpha\equiv(-a_ia^i)^{1/2}$ in the near proximity of the
semi-infinite rod ($r\ll |z|$) we can neglect the contribution to $\psi$
coming from the finite rod $z_1z_2$, since its presence
cannot give rise to a divergence of $\alpha$.
Then
\begin{equation}\label{4.3}
e^{2\psi}\approx(r^2+z^2)^{\frac 12}+z, \
e^{2\psi-2\gamma}\approx 2(r^2+z^2)^{\frac 12},
\end{equation}
and the relevant term of the squared norm of the acceleration
defined above reads
\begin{equation}\label{4.4}
{\alpha}^2=e^{2\psi-2\gamma}
\left[\left(\frac{\partial\psi}{\partial z}\right)^2
+\left(\frac{\partial\psi}{\partial r}\right)^2\right]
\approx\frac{1}{2\left[(r^2+z^2)^{\frac 12}+z\right]}.
\end{equation}
When $z$ is negative $\alpha$ diverges when the
limit $r\rightarrow 0$ is taken, i.e. when the world line of
absolute rest is drawn closer and closer to the horizon
produced by the semiinfinite rod.\par
In the Kruskal manifold, a similar intrinsic singularity
occurs when considering the norm of the four acceleration
along a line of absolute rest located, in the left and right quadrants,
at positions closer and closer
to the bifurcate horizon, possibly to warn that it is not
a good idea to envisage joining manifolds of
different ``relativity content'', and that Schwarzschild's
original manifold \cite{Schwarzschild1916} is all what
is allowed to provide a model for the spherically
symmetric gravitational field of a particle.\par
The same occurrence happens with the ``boost-rotation
symmetric'' manifolds. In this case too, one inclines
to think that the singular behaviour of $\alpha$ under
analogous circumstances is again there to
spell the same kind of warning.
\section{Conclusion}
The standard view about the vacuum C-metric \cite{Bonnor1983} and its relatives,
as discussed e.g. in
\cite{Bondi1957}, \cite{BS1964}, \cite{IK1964},
\cite{Bonnor1966}, \cite{Bicak1968,BHS1983,BS1984},
\cite{Bonnor1988}, \cite{BS1989}, assumes that the singularities
representing the nonspinning masses of these vacuum solutions exhibit a
uniformly accelerating motion relative to an inertial frame at infinity.
This interpretation is problematic, since it relies on approximate,
asymptotic group symmetries of the corresponding manifolds, while
the exact Killing group symmetry that prevails everywhere in the
submanifolds where the world lines of the masses are located
shows that the nonspinning masses are at rest
with respect to the latter, intrinsically static submanifolds in the invariant,
absolute sense explained in Section 3. Moreover the submanifolds
that contain the world lines of the masses are joined to the
remaining parts of the manifolds at hypersurfaces that are
singular in the invariant, local, intrinsic sense expounded in Section 4.

\appendix
\section{Weyl's method of solution}\label{A}
In the static, axially symmetric case, despite the nonlinear structure
of Einstein's field equations, Weyl succeeded in reducing the problem
to quadratures through the introduction of his ``canonical cylindrical
coordinates''. Let $x^0=t$ be the time coordinate, while $x^1=z$, $x^2=r$ are the
coordinates in a meridian half-plane, and $x^3=\varphi$ is the
azimuth of such a half-plane; then the line element of a static,
axially symmetric field {\it in vacuo} can be tentatively
written as:
\begin{equation}\label{A1}
\dd s^2=e^{2\psi}\dd t^2-\dd\sigma^2,\;e^{2\psi}\dd\sigma^2
=r^2\dd\varphi^2+e^{2\gamma}(\dd r^2+\dd z^2);
\end{equation}
the two functions $\psi$ and $\gamma$ depend only on $z$ and $r$.
Remarkably enough, in the ``Bildraum'' introduced by Weyl $\psi$
fulfils the potential equation
\begin{equation}\label{A2}
\Delta\psi=\frac{1}{r}\left\{\frac{\partial(r\psi_z)}
{\partial z}
+\frac{\partial(r\psi_r)}{\partial r}\right\}=0
\end{equation}
($\psi_z$, $\psi_r$ are the derivatives with respect to $z$ and to
$r$ respectively), while $\gamma$ is obtained by solving the system
\begin{equation}\label{A3}
\gamma_z=2r\psi_z\psi_r,\;\gamma_r=r(\psi^2_r-\psi^2_z);
\end{equation}
due to the potential equation (\ref{A2})
\begin{equation}\label{A4}
\dd\gamma=2r\psi_z\psi_r\dd z+r(\psi^2_r-\psi^2_z)\dd r
\end{equation}
happens to be an exact differential.

\newpage

\bibliographystyle{amsplain}

\end{document}